\documentclass[12pt]{amsart}

\usepackage[margin=1in]{geometry}
\usepackage[lofdepth,lotdepth,caption=false]{subfig}
\usepackage{fancyhdr}
\usepackage{hyperref}
\usepackage{faktor}
\usepackage{amsmath, amssymb, graphicx}
\usepackage{mathtools}
\usepackage{xspace}
\usepackage{braket}
\usepackage{yfonts}
\usepackage{multicol}
\usepackage{multirow}
\usepackage{color}
\usepackage{setspace}
\usepackage{enumitem}
\usepackage{pst-node}
\usepackage{tikz-cd}
\usepackage{tikz}
\usetikzlibrary{decorations.markings}
\usepackage{slashed}
\usetikzlibrary{calc}
\usepackage[mathscr]{euscript}
\usepackage[numbers]{natbib}
\usepackage[capitalize,nameinlink]{cleveref}

\newcommand{\ra}{\rightarrow}
\newcommand{\pr}{\prime}

\newcommand{\C}{\mathbb{C}}

\newcommand{\R}{\mathbb{R}}

\DeclareMathAlphabet{\mathpzc}{OT1}{pzc}{m}{it}

\theoremstyle{plain}

\theoremstyle{definition}
\newtheorem{dfn}{Definition}[section]
\theoremstyle{remark}
\newtheorem{rmk}{Remark}[section]

\newtheorem*{note}{Note}

\begin{document}

\title{The Universe from a Single Particle III}
\maketitle

\begin{center}
  \normalsize
  Michael Freedman \footnote{michaelf@microsoft.com}\textsuperscript{,\hyperref[1]{*}},
  Modjtaba Shokrian Zini \footnote{mshokrianzini@pitp.ca} \textsuperscript{,\hyperref[2]{$\dagger$},\hyperref[3]{$\bullet$}}
 \par   \bigskip
\end{center}

\begin{abstract}
    In parts I \cite{freedman2021universe} and II \cite{freedman2021universeII} we studied how metrics $g_{ij}$ on $\mathfrak{su}(n)$ may spontaneously break symmetry and \emph{crystallize} into a form which is kaq, \emph{knows about qubits}. We did this for $n=2^N$ and then away from powers of 2. Here we address the Fermionic version and find kam metrics, these \emph{know about Majoranas}. That is, there is a basis of principal axes $\{H_k\}$ of which is of homogeneous Majorana degree. In part I, we searched unsuccessfully for functional minima representing crystallized metrics exhibiting the Brown-Susskind penalty schedule, motivated by their study of black hole scrambling time. Here, by segueing to the Fermionic setting we find, to good approximation, kam metrics adhering to this schedule on both $\mathfrak{su}(4)$ and $\mathfrak{su}(8)$. Thus, with this preliminary finding, our toy model exhibits two of the three features required for the spontaneous emergence of spatial structure: 1. localized degrees of freedom and, 2. a preference for low body-number (or low Majorana number) interactions. The final feature, 3. constraints on who may interact with whom, i.e. a neighborhood structure, must await an effective analytic technique, being entirely beyond what we can approach with classical numerics.
\end{abstract}

\tableofcontents

\section{Introduction}
The most symmetrical metric on $\mathfrak{su}(n)$ is the Killing form $\langle H, H \rangle = -2n\operatorname{tr}(H^2)$. This metric is adjoint-invariant and induces a left-invariant metric on $\operatorname{SU}(n)$ of diameter $\pi$. Less symmetric metrics $g_{ij}$ on $\mathfrak{su}(n)$ are motivated by \emph{quantum compiling} and black hole physics. Both contexts suggest metrics diagonal in the Pauli-word basis \cite{nielsen2005geometric,nielsen2006quantum,dowling2008geometry,brown2017quantum,brown2018second}. In this paper $n = 2^N$ and a \emph{Pauli word} is an $N$-fold tensor product with a 1, $X$, $Y$, or $Z$ in each slot, e.g.\footnote{$\sqrt{-1}$ is chosen to make the word skew-Hermitian.}\ $\sqrt{-1} \ 1 \otimes X \otimes 1 \otimes Y \otimes X \otimes 1 \otimes 1 \otimes Z$ is a word of \emph{weight} $w = 4$ (4 letters) in $\mathfrak{su}(2^8)$. Low weight directions are both the \emph{practical} directions along which to evolve a quantum state in a quantum computer and according to the SYK \cite{trunin2020pedagogical} model (in its Bosonic version), the principal directions of black hole evolution \cite{brown2017quantum,brown2018second}. These constraints direct our attention to the Brown-Susskind exponential penalty metrics:
\begin{equation}\label{eq:pen-met}
    g_{ij} = \delta_{ij} e^{\text{const. weight}(i)}.
\end{equation}

In parts I and II \cite{freedman2021universe,freedman2021universeII}, we studied functionals on the space of metrics, recalled briefly in section 2. The chief finding was that a surprising number of local minima create and respect a tensor product structure on the underlying Hilbert space. We called these metrics kaq for \emph{knows about qubits} and the process of falling into such a minimum a \emph{metric crystallization} in analogy to the formation of crystals through spontaneous symmetry breaking. However, in part I, we did not find crystallized metrics of Brown-Susskind type \cref{eq:pen-met}, either from random initialization or from seeds of exactly that form. Indeed, initializing in metrics obeying \cref{eq:pen-met} always led through gradient descent to unrelated minima.

But just as the SYK model appears most useful in Fermionic form where 2-body interactions appear as
\begin{equation}\label{eq:Fermionic-pm}
    H_{\mathrm{SYK}} = \sum_{i,j,k,l} J_{ijkl} \gamma_i \gamma_j \gamma_k \gamma_l,
\end{equation}
a Fermionic analogy of Brown-Susskind metric would have a first (random) term like \cref{eq:Fermionic-pm} and continue with higher order interactions with exponentially decaying coefficients. We observe local minima on both $\mathfrak{su}(4)$ and $\mathfrak{su}(8)$ of this form. It is true that finding Fermionic Brown-Susskind minima requires careful seeding, but similar care on the qubit side failed to reach metrics of the form \cref{eq:pen-met}. This suggests that Fermionic Brown-Susskind metrics arise from spontaneous symmetry breaking, and makes them natural candidates for $\operatorname{H}_{\text{initial}}$, the initial Hamiltonian of the universe, at least with the toy model under discussion. In \cite{brown2021effective}, the concept of a universal \textit{critical} metric is developed axiomatically. Our data here may reflect on both $\mathfrak{su}(4)$ and $\mathfrak{su}(8)$ critical metrics with a roughly exponential structure. In both cases we see a brief dip in eigenvalues prior to their exponential growth. What stands out is the preference for low degree Majorana monomials rather than the precise exponential form.

We have already described Pauli word basis, here in more detail is the Majorana basis for $\operatorname{SU}(2^N)$. Just as in \cite{freedman2021universe,freedman2021universeII} where there was a variable isomorphism $J: (\C^2)^{\otimes N} \xrightarrow{\cong} \C^{2^N}$ (and the induced $j: \operatorname{Her}(2)^{\otimes N} \xrightarrow{\cong} \operatorname{Her}(2^N)$ in the background), here there is also the same choice. We may conjugate by any $\operatorname{U} \in \operatorname{SU}(2^N)$ to transform the coordinates on $\mathfrak{su}(2^N)$. To have a precise $\ast$-isomorphism, we need to identify the complexified Real Clifford (Majorana) algebra $\operatorname{Cliff}_\R(2^N) \otimes_\R \C$, where
\[
    \operatorname{Cliff}_{\R}(2^N) = \R[\gamma_1, \dots, \gamma_N],\ \{\gamma_i,\gamma_j\} = 2\delta_{ij},\ \gamma_i^\ast = \gamma_i,
\]
with $\mathfrak{u}(2^N) \otimes_\R \C$ represented as follows:
\begin{equation}
\begin{split}
    & \gamma_1 \ra  1 \cdots 1 X,\ \gamma_2 \ra  1 \cdots 1 Y,\ \gamma_3 \ra  1 \cdots 1 XZ, \gamma_4 \ra  1 \cdots 1 YZ \\
    & \gamma_5 \ra  1 \cdots 1 XZZ,\ \gamma_6 \ra  1 \cdots 1 YZZ,\ \gamma_7 \ra  1 \cdots 1XZZZ,\ \gamma_8 \ra 1 \cdots 1YZZZ, \text{etc.}
\end{split}
\end{equation}

\begin{dfn}\label{kamdfn}
    We say a metric $g_{ij}$ on $\mathfrak{su}(2^N)$ knows about Majoranas if it is \emph{not} proportional to the ad-invariant Killing metric (in which case our condition is trivial) yet possesses an orthonormal basis $\{H_k\}$ of principal axes (w.r.t. the Killing form) so that for all $1 \leq k \leq 4^N-1$, $H_k$ is of homogeneous Majorana degree, i.e.\ a polynomial in the $\gamma$'s (under the above identification) with a constant number of $\gamma$'s in each term.
\end{dfn}

\begin{note}
    In many applications, operators of odd Majorana degree are not physical because their application would violate the Fermionic parity super-selection rule. There is no similar issue here, we are simply using $\Gamma$-matrices to write out a basis for a Lie algebra. Both even and odd degree Majorana operators (after judicious insertion of powers of $\sqrt{-1}$) are legitimate basis elements of $\mathfrak{su}(2^N)$.
\end{note}

Although we hope to study metric crystallization analytically in the future (e.g.\ the link \cite{hoppe1989diffeomorphism} between $\operatorname{SU}(n)$ and 2D Hamiltonian system might provide an avenue), this paper is numerical, and we must allow some tolerance around the ideal definition. If, from a random seed, we find a metric $g_{ij}$ so that a related $g_{i^\pr j^\pr}$ has 80\% or more of a principal axes basis each with at least 95\% of its $L^2$-norm concentrated in a single homogeneous degree subspace we consider the minimum to be kam. In \cite{freedman2021universe}, we did dimension counting to demonstrate the rarity of kaq metrics. To make a similar case for the rarity of kam or kaq metrics, up to an exponential tolerance, one ideally would estimate the phase space volume satisfying our acceptance rule. Although undoubtedly tiny, a rigorous estimate would require a feat of algebraic geometry. We instead adopt an expedient. To complement the roughly 100 gradient descents from fully random or random Majorana diagonal seeds, carried out for this study, we randomly generate a similar number of metrics $g_{ij}$ to be used as a \emph{control group}. These $g_{ij}$ are random except for the specification of the principal axis degeneracy pattern which we chose to mimic Majorana degree degeneracies, e.g.: (4, 6, 4, 1) in the case of $\mathfrak{su}(4)$ and (6, 15, 20, 15, 6, 1) in the case of $\mathfrak{su}(8)$. Then we search the possible conjugate metrics $g_{i^\pr j^\pr}$, as above, for \emph{accidental} kam structure. In no case was our experimental criterion close to being met.\footnote{Having found this a reassuring sanity check, in \cref{randomkaqs} we return to part I \cite{freedman2021universe} and apply a similar check. Again, we find no \emph{accidental} kaq metrics.}

Before giving the details of our search methods and the results, we should explain that search is done from three qualitatively different types of initial metrics $g_{ij}$ \cite[Refer to Section 3 for a fully detailed list]{freedman2021universe}:
\begin{enumerate}
    \item Fully random seed (\cite[\textbf{GenPerturbId}]{freedman2021universe}). Here the metric is selected from a Gaussian centered at the ad-invariant Killing metric, which we call $1_n$, since it appears as the identity when written in either a Pauli-word-weight, or a Majorana-degree basis. For reasons of numerical stability we choose a Gaussian of small variance and generally use a slow learning rate to avoid inadvertently jumping over nearby local minima .

    \item Random diagonal (\cite[\textbf{DiagPerturbId}]{freedman2021universe}). Here the diagonal entries are iid Gaussian distributed (then normalized), the off-diagonal entries are 0. Our rigorous analysis from \cite[See theorems in Appendix B]{freedman2021universe} shows that for the class of functions we treat, gradient flow preserves the diagonal condition. In our numerics, we enforce this conservation law exactly by defining the metric with $\dim \mathfrak{su}(2^N) = 4^N - 1$ learnable parameters on the diagonal.

    \item Random diagonal with batched eigenvectors which we call \textbf{BatchedDiagPerturbId}, batched according to homogeneous Majorana degree. \cite[Thm. B2]{freedman2021universe} demonstrated that batched groups of eigenvectors will \emph{stay} batched\footnote{Batches may merge or have their eigenvalues cross through each other, but cannot split.} under gradient flow, provided the eigendirections are faithfully permuted (modulo powers of $i$) by a group symmetry. The relevant symmetry is the symmetric group $\operatorname{S}(2N)$ acting on $\{\gamma_i\}_{i=1}^{2N}$. \label{BatchedDiagPerturbId}
\end{enumerate}

In case (1), $g_{ij}$ has no initial structure so emergence of kaq or kam metrics is most surprising. Cases (2) and (3) increasingly ``stack the deck'' making it easier to locate local minima of interest. The functionals we study (section 2) are on spaces of metrics having hundreds of dimensions (2015 dimension for $\mathfrak{su}(8)$) and many local minima, and require such initialization to fully explore. In cases (2) and (3), what we are looking for is strictly \emph{unforced} behavior. In the case of (2), this would be the formation of eigenvalue degeneracies associated to Majorana degree and perhaps sub-Lie-algebra structures (although these were not found from Majorana initializations). In case (3), the independent variable is the eigenvalues or $\text{lengths}^2$ of principal axes. The finding highlighted above of Fermionic Brown-Susskind metrics was the result of a type (3) initialization.

It is natural to inquire if this finding could be due to chance. Although we do not have enough data for a careful statistical study, for comparison, for each of the 7 functionals analyzed, we generated 10,000 random functions $f$ of $\{1,\dots,6\}$ to represent possible eigenvalues at the local minimum for the batches of degree $d$ eigenvalues, $1 \leq d \leq 6$, corresponding to the Majorana basis for $\mathfrak{su}(8)$. The value was selected uniformly between the smallest and largest eigenvalues seen in our actual runs. For each function the loss for the best $L^2$-fit to the exponential form was evaluated. and compared to the mean loss of our actual runs. One functional particularly stands out as always giving approximately a Fermionic Brown-Susskind structure in its local minima, while others struggle to do so. We refer to \cref{nullhyporandomfunctions} for more details.

\section{Review of functional}

We review the perturbed Gaussian integral (inspired from \cite{bar1995perturbative}) used to define the functionals in \cite{freedman2021universe}. Let
\begin{align}\label{theintegral}
    & F_k := \int_{\vec{x} \in \R^{3(4^n-1)}}\ \operatorname{d}\vec{x}\ e^{ik(G_{IJ}x^I x^J + c_{ijk}y_1^i y_2^j y_3^k)}
\end{align}
where $x = (y_1,y_2,y_3)$ with $y_o \in \R^{4^n-1}, o \in \{1,2,3\}$, and for $I = (i,o)$, $x^I = y_{o}^i \in \mathbb{R}$, and $G_{IJ}x^I x^J = g_{ij}y_1^i y_1^j + g_{ij} y_2^i y_2^j + g_{ij} y_3^i y_3^j$, i.e. $G = \begin{pmatrix}
    g & 0 & 0 \\
    0 & g & 0 \\
    0 & 0 & g
\end{pmatrix}$. The structure constants $c_{ij}^k$ of the Lie algebra are
\begin{equation}\label{c_ijkdfn}
    [y_i, y_j] = c_{ij}^k y_k \text{ and } c_{ijk} = c_{ij}^{k^\pr} g_{k^\pr k}.
\end{equation}
The real and imaginary part of $F_k$ are of interest:
\begin{align}
    f_{k,1} = \operatorname{Re}(F_k),\ f_{k,2} = \operatorname{Im}(F_k).
\end{align}

From the two functionals above, We derive two functionals called $F_{26}(c,g,k)$ and $F_{24}(c,g,k)$. The subscripts denote how far the perturbative expansion is computed. For $F_{26}$, we compute the 2 and 6 vertex diagrams, and for $F_{24}$, the 2 and 4 vertex diagrams. The details are discussed in The in \cite[Section 2.1]{freedman2021universe} and also reviewed in \cite[Section 3]{freedman2021universeII}.

To find the local minima of these functionals, we obviously need to fix a volume for $g$, i.e. set $\det g = 1$. To enforce this condition, it is numerically more stable to take a Lagrangian approach instead of normalizing by $\det(g) = 1$ \cite{freedman2021universe}:
\begin{align}\label{L_24}
    L_{24}(c,g,k) = r_1^{-1}F_{24}(c,g,k) + r_2(\det(g)-1)^2, \\\label{L_26}
    L_{26}(c,g,k) = r_1^{-1}F_{26}(c,g,k) + r_2(\det(g)-1)^2,
\end{align}
where $r_1 \ge 1, r_2 >>1$. Gradient descent on these two functionals yield the solution metrics we analyze for \textbf{kam}ness. Our numerics always work through the Feynman-Penrose asymptotic expansion; the integral itself is oscillating and approaching it through Riemann sums would not be fruitful. As mentioned before, these solutions generally have highly degenerate eigenspaces. 
\begin{dfn}[$\text{\cite[Definition 2.1]{freedman2021universe}}$]\label{degpatdfn}
The \textit{degeneracy pattern} $(d_1,\ldots,d_t)$ is a tuple describing the dimensions of the eigenspaces ordered by increasing eigenvalues, i.e. from \textit{easier} to \textit{harder} directions.
\end{dfn}
For $N=2,3$ we refer to \cite[Tables 1-2 (\textbf{GenPerturbId})]{freedman2021universe} for the values chosen for $k,r_1,r_2$. We simply note that for $N=2$ we always choose $k=100,200$ and for $N=3$, we choose $k=500,1000$. Here, as in Chern-Simons theory \cite{bar1995perturbative}, $\frac{1}{k}$ serves as an expansion parameter as it controls the relative weights of the quadratic and cubic terms. The ability to pick $\frac{1}{k}$ small, stabilizes the numerics.

\section{Know-about-Majoranas search}
We reexamine the solutions found in \cite{freedman2021universe} for \textbf{kam}ness. Each solution $g$ comes with eigenvectors $\{iH_1,\ldots,iH_{n^2-1}\}$, where $H_j$ are hermitian $n \times n$ matrices with $l_2$ norm normalized (recall $n=2^N$).

\subsection{Kam loss function}~
\\
\indent We want to design a loss function, which global minimum is 0 if and only if the solution $g$ is \textbf{kam}.

\subsubsection{Identifying the parameters of the loss function} Following \cref{kamdfn}, there are two sources for the parameters of such a loss function. This is identical to \textbf{kaq} loss function defined in \cite[Section 3.2]{freedman2021universeII}). The first set of parameters describe the conjugation of the eigenbasis by some $U\in U(n)$. Next, note that the choice of the basis of each degenerate eigenbasis is not unique, and so a degenerate eigenspace of degree $d$ can afford an independent change of basis, leading to the second source of parameters of our loss function, which describe an orthogonal matrix $V \in O(d)$. The total number of parameters is $n^2 + \sum_{i=1}^{t} (d_i^2-d_i)/2$ where $(d_1,\ldots,d_t)$ is the degeneracy pattern of $g$. We use $\theta$ to denote all these parameters. ~

\subsubsection{Computing projection to homogeneous Majorana spaces} After the above two transformations on the eigenstates, abusing the notation, let the new eigenstates be $\{iH_1,\ldots,iH_{n^2-1}\}$. Then we compute $v_{p,q}$ which is the \textbf{squared} projection norm of $H_{p}, 1\le p\le n^2-1$, to the homogeneous Majorana space of degree $q$. We compute this as we would compute it for a vector projection to a subspace given by its orthonormal basis. Here, the subspace is given by the orthonormal basis $\{\gamma_{i_1}\ldots\gamma_{i_q}\}_{1\le i_1 < \ldots <i_q \le 2N}$.

\subsubsection{Formula for the loss} For each $p$, since the homogeneous Majorana spaces of degree $q$ span the whole hermitian matrix space, we have $\sum_q v_{p,q} = ||H_p||_2^2 = 1$. Clearly, we would like one of the projection norms to be one, and thus the rest to be zero. To have a loss function $\mathcal{L}_\theta(g)$ with minimum described by such a configuration, we can simply define:

\begin{align}
    \mathcal{L}_\theta(g) = \sum_{p=1}^{n^2-1} L_\theta(H_p), \ \text{where} \\
    L_\theta(H_p) = \prod_{q=1}^{2N} (1-v_{p,q}).
\end{align}
It is not hard to see that $\mathcal{L}_\theta(g) = 0 $ for some parameters $\theta$ iff $\forall p: L_\theta(H_p)=0$ iff $\forall p \exists q: v_{p,q} = 1$, i.e. $g$ is \textbf{kam}. There are other possible designs for $\mathcal{L}_\theta$, like the sum of $(1-v_{p,q})^2$ for $L_\theta(H_p)$, which we note, changes the global minimum of $\mathcal{L}_\theta$ when $g$ is \textbf{kam}. We tried these other formulae and they did not give us any other \textbf{kam} solutions. 

\subsection{Kam solutions tables}~
\\
\indent We make this section very similar to \cite[Section 4.2]{freedman2021universeII}, where the local minima found in \cite{freedman2021universe} for $N=2,3$ through a \textbf{GenPerturbId} search are listed by by their degeneracy patterns and their \textbf{kam}ness.

\subsubsection{Remarks on the results}\label{rmkpresentation}
\begin{enumerate}
    \item When a solution is declared to be \textbf{kam}, the value of $\mathcal{L}_\theta$ is very low, smaller than $1e-3$, and as a result all $L(H_p)$ are smaller than $1e-4$. On the other hand, in our experience, there has been a clear line between \textbf{kam} and non-\textbf{kam} solutions, where $\mathcal{L}_\theta$ is at least 1 (or in most cases, esp. for $N=3$, much larger than 1).
    \item Within the description and captions, we will use $(d_i)$ for the degeneracy pattern (\cref{degpatdfn}) and thus, $d_i$ refers to the dimension of an eigenspace.
    \item In some of the tables, we give some explanation on the solutions and their \textbf{kam}ness. Sometimes no \textbf{kam}s are found, esp. for $N=3$, in which case we still show how close to being \textbf{kam} some solutions were. This should be compared to our random simulations results in the next section, showing that our solutions are still quite rare when considering how close to \textbf{kam} they are.
    \item ``$(d_1,\ldots,d_t): x/y$'' means $x$ solutions out of the $y$ solutions with pattern $(d_1,\ldots,d_t)$ are \textbf{kam}.
    \item Some tables only show solutions for a single value of $k$. This is when the lower value gave solutions very close to identity (see \cite[Remark 3.4]{freedman2021universeII}).
\end{enumerate}

\begin{table}[h]
    \centering
    \begin{center}
    \begin{tabular}{ |p{8cm}|p{8cm}| } 
    \hline
    $k=100$ & $k=200$ \\
    \hline
    $(10,5)$: 3/3  \newline $(1,4,8,2)$: 4/14  &  No \textbf{kam} found. However the pattern $(1,1,2,4,2,1,2,2)$ with $\mathcal{L}(g)\sim 0.61$ which was the result of which was the result of four eigenstate having $L(H_p) \sim 0.153$ and the rest being (very close to) zero.   \\
    \hline 
    \end{tabular}
    \end{center}
    \caption{\textbf{Kam}ness for $L_{24}$ on $\mathfrak{su}(4)$.}
    \label{l24results4}
\end{table}
\begin{table}[h]
    \centering
    \begin{center}
    \begin{tabular}{ |p{8cm}|p{8cm}|} 
    \hline
    \multicolumn{2}{|c|}{$k=1000$} \\
    \hline
    % sec1,last1
    \multicolumn{2}{|p{16cm}|}{$(1,16,1,6,2,16,16,1,2,2)$ : 0/13. However, for some instances, we had $\mathcal{L}(g)\sim 3.5$ with $14$ many $L(H_p) \sim 0.25$ and the rest close to zero.} \\
    \hline 
    \end{tabular}
    \end{center}
    \caption{\textbf{Kam}ness for $L_{24}$ on $\mathfrak{su}(8)$.}
    \label{l24results8}
\end{table}
\begin{table}[h]
    \centering
    \begin{center}
    \begin{tabular}{ |p{8cm}|p{8cm}| } 
    \hline
    $k=100$ & $k=200$ \\
    \hline 
    % 10 - 1,1 or sth else - 10 - 8 - 1,1 or sthelse - 1,1
    $(10,5)$: 3/3  \newline $(3,1,1,8,2)$: 1/11 &  $(10,5)$: 1/1  \newline $(8,6,1)$: 0/1  \newline $(3,1,4,2,1,4)$: 0/7  \newline $(1,3,1,4,4,2)$: 0/6\\
    \hline 
    \end{tabular}
    \end{center}
    \caption{\textbf{Kam}ness for $L_{26}$ on $\mathfrak{su}(4)$.}
    \label{l26results4}
\end{table}
\begin{table}[h]
    \centering
    \begin{center}
    \begin{tabular}{ |p{8cm}|p{8cm}| } 
    \hline
    $k=500$ & $k=1000$\\
    \hline
    No \textbf{kam} found. Nevertheless, for the pattern $(10,15,1,32,5)$, best instances had $\mathcal{L}(g) \sim 1.5$, with six many $L(H_p) \sim 0.25$ and the rest close to zero. & Similar to $k=500$: Best instances of the pattern $(10,15,1,32,5)$ had $\mathcal{L}(g) \sim 1.5$, with six many $L(H_p) \sim 0.25$ and the rest close to zero.\\
    \hline 
    \end{tabular}
    \end{center}
    \caption{\textbf{Kam}ness for $L_{26}$ on $\mathfrak{su}(8)$.}
    \label{l26results8}
\end{table}

\begin{rmk}
Compared to the \textbf{kaq}ness results in \cite{freedman2021universeII}, we see that it is much easier to find a \textbf{kaq} solution than \textbf{kam}. Nevertheless, both are rare as shown in the next section.
\end{rmk}

\subsection{Null Hypothesis: Searching for random \textbf{kam}s/\textbf{kaq}s}
\subsubsection{Random \textbf{kam}s} As discussed before, we take 100 randomly generated metrics $g_{ij}$ with a Majorana degree degeneracy pattern. We do so by first randomly generating a diagonal metric with such a pattern, and conjugate it by a random orthogonal matrix. As a result the random metric has the Majorana degree degeneracy pattern, but whether it is \textbf{kam} or not depends on the random orthogonal matrix. After running gradient descent for each 100 randomly generated metric, we found no instance of \textbf{kam}:
\begin{itemize}
    \item For $\mathfrak{su}(4)$: The vast majority 97/100 had $\mathcal{L}(g)>2$, and three had loss $\sim 1.5$. Even for those three random metrics, none of the $L(H_p)$ were smaller than $0.002$, meaning no eigenstate met our criteria ($1e-4$) to be a homogeneous Majorana degree subspace in the minimum for $\mathcal{L}(g)$. This stands in stark contrast with the non-\textbf{kam} pattern found in \cref{l24results4} for $k=200$.
    \item For $\mathfrak{su}(8)$: The lowest loss was $\mathcal{L}(g) \sim 12$, with the least $L(H_p)$ being $0.01$. Again, this is in contrast with the non-\textbf{kam} pattern found in \cref{l26results8} for $k=200$.
\end{itemize}

\subsubsection{Random \textbf{kaq}s}\label{randomkaqs} Similarly, we do random simulations to search for \textbf{kaq} patterns. For notations, we refer to \cite[Section 3.2]{freedman2021universeII}:
\begin{itemize}
    \item For $\mathfrak{su}(4)$: Degeneracy pattern is (6, 9). For the vast majority $\mathcal{L}_\textbf{kaq}(g) > 2$ with a few $\sim 1$. Lowest entropy $s_{j}$ was $0.002$, meaning no eigenstate $H_j$ could be factored to a tensor product (we have a $1e-4$ criteria, similar to $L(H_p)$ for \textbf{kam}ness). This is also in contrast with non\textbf{kaq} solutions found in \cite{freedman2021universeII}, which had the vast majority of their eigenstates factorized.
    \item For $\mathfrak{su}(8)$:  Note that we searched for \textbf{partial-kaq}, i.e. a $\C^4 \otimes \C^2$  decomposition, which is more likely to occur than a \textbf{kaq} decomposition $\C^2 \otimes \C^2 \otimes \C^2$. The typical loss was $\sim 35$, but if two eigenspaces out of the three (9, 27, 27) were to be merged, thus giving more degrees of freedom to find a \textbf{kaq} configuration, then loss dropped to 12, with no $H_p$ being factorized. Again, this is in contrast with the non-\textbf{kaq} pattern found in \cite{freedman2021universeII}.
\end{itemize}

\section{Fermionic Brown-Susskind metrics search}

\subsection{The search setting}~
\\
\indent To search for Fermionic Brown-Susskind (FBS) metrics, as discussed previously in \cref{BatchedDiagPerturbId}, we use the \textbf{BatchedDiagPerturbId} method. Furthermore, we make our batched diagonal initialization on a random FBS metric determined by a weight $w$, i.e. $g_{ij}=\delta_{ij}w^{\text{weight(i)}}$ where $\text{weight}(i)$ is the Majorana degree of basis element $i$. Although we proved in \cite{freedman2021universe} that the gradient descent preserves the degeneracy pattern, in no way is it bound to preserve the FBS nature of the metric and neither the ascending order of the eigenvalues as illustrated later. Indeed, the parameters of the gradient descent are the weights $w_d$ for degree $d$ Majorana monomials, and not $w$ (although initially $w_d = w^d$). Thus in the solutions obtained, some exhibit approximately an FBS structure, while some do not. Of those that do, notice that the data shows a small dip prior to the exponential rise, we have speculated that this could be a signature of the critical metric proposed in \cite{brown2021effective}.

\subsection{Data, charts and graphs}
\subsubsection{FBS solutions graph}\label{solsgraphsection} 
\begin{figure}[h]
    \centering
    \includegraphics[scale = 0.5]{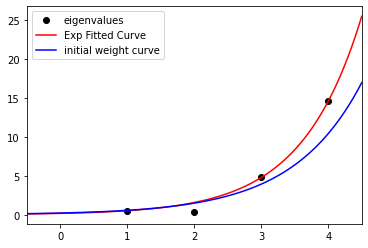}
    \caption{$L_{24}(k=200)$ solution with exponential fit loss $\sim 0.34$. For this functional on $\mathfrak{su}(4)$, the 12 searches we did all turned up the same local minima. To see how rare this single exponential fit loss is, we can randomly simulate 10,000 functions similar to \cref{nullhyporandomfunctions}, but with the additional constraints that $0.85max \le \max_{1\le j\le 2N} f(j) \le max$ and that the exponent $b>0$, then the loss of this solution would be higher than only 0.84\% of the random losses. Note these additional constraints make the comparison fairer, as higher maximum eigenvalue generally means higher $L^2$-loss (since the loss is not scale-invariant), and requiring $b>0$ ensures we look at samples that have an overall increasing set of eigenvalues, as is ours.}
    \label{expsu4}
\end{figure}

\begin{figure}[h]
    \centering
    \includegraphics[scale = 0.5]{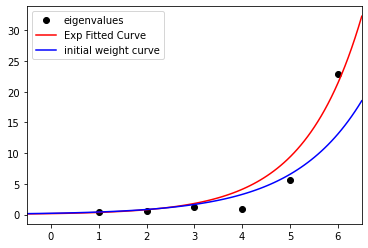}
    \includegraphics[scale = 0.5]{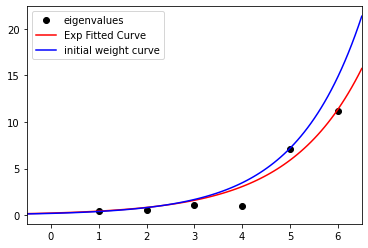}
    \caption{Two solutions for $L_{24}(k=1000)$ with exponential fit loss $\sim 4.55, 1.02$ (left, right respectively). These are found among the 12 solutions for this functional, of which there are 5 distinct solutions. Unlike in \cref{expsu4}, this functional had different local minima. If we were to randomly simulate 10,000 functions similar to \cref{nullhyporandomfunctions}, but with the additional constraints that $0.85max \le \max_{1\le j\le 2N} f(j) \le max$ and that the exponent $b>0$, once for the left solution $max \sim 22.8$ and once for the right solution $max \sim 11.16$, then the loss of the left and right solutions would be higher than 3.95\%, 0.97\%  of the random losses.}
    \label{expsu8}
\end{figure}

In \cref{expsu4} and \cref{expsu8}, we plot the solutions that exhibited an approximate FBS structure. We favor these plots since they have some of the highest maximum eigenvalue compared to other plots, and yet achieve relatively low loss. In addition to scatter plotting the eigenvalues, we have plotted the best exponential fit of the form $e^{b(x-\frac{3.4^{N-1}N}{(4^N-1)})}$ where $b$ is the only parameter of the fit, and $-\frac{3.4^{N-1}N}{(4^N-1)}$ is to enforce $\det = 1$. Furthermore, the initial exponential curve (given by $w^{d-\frac{3.4^{N-1}N}{(4^N-1)}}$) that gives the initial metric. The $y$ coordinate determines the eigenvalue, while the $x$ coordinate determines the degree of the Majorana monomial.

\subsubsection{Not-FBS solutions graph} Below, in \cref{badexpsu4} and \cref{badexpsu8}, we also reproduce the solutions that were far away from an exponential fit.

\begin{figure}[h]
    \centering
    \includegraphics[scale = 0.5]{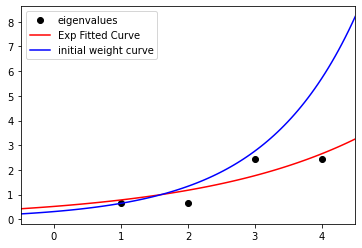}
    \includegraphics[scale = 0.5]{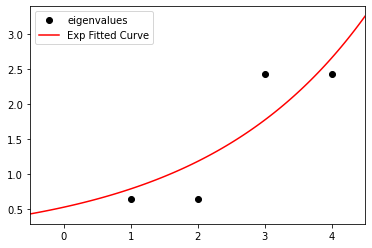}
    \includegraphics[scale = 0.5]{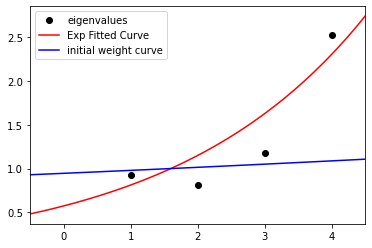}
    \caption{Two bad fits from $L_{26}(k=100)$ (top left and right) and $L_{24}(k=100)$ (bottom). The top left and right graphs are for the same solution, and the right graph shows only the eigenvalues and the fitted curve in order to better illustrate the bad fit. As a comparison, the fit loss of the top graph was larger than 70\% of the respective random losses found in \cref{nullhyporandomfunctions}.}
    \label{badexpsu4}
\end{figure}

\begin{figure}[h]
    \centering
    \includegraphics[scale = 0.5]{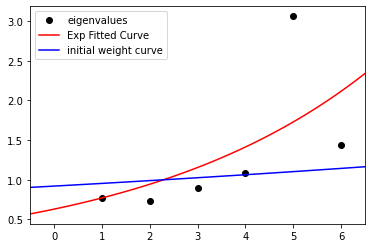}
    \includegraphics[scale = 0.5]{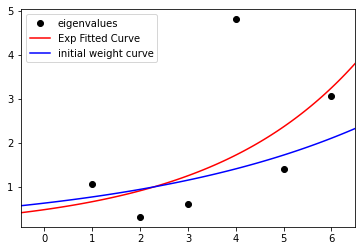}
    \includegraphics[scale = 0.5]{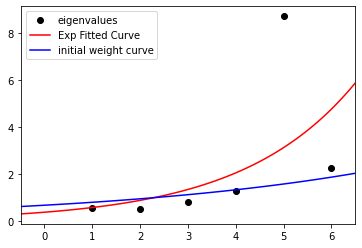}
    \caption{Top left and right are two different solutions from $L_{26}(k=500)$ and $L_{26}(k=1000)$, while the bottom is a solution from $L_{24}(k=1000)$. As a comparison, the top left fit loss was larger than 44\% the respective random losses found in \cref{nullhyporandomfunctions}.}
    \label{badexpsu8}
\end{figure} 
\newpage

\subsection{Null hypothesis: Searching for random FBS}\label{nullhyporandomfunctions}~
\\
\indent To see how \textit{rare} the exponential fit loss of the aforementioned graphs are, we review the process mentioned in the introduction: we generate 10,000 random functions $f$ of $\{1,\dots,2N\}$ representing possible eigenvalues at the local minimum for the batches of degree $d$ eigenvalues, $1 \leq d \leq 2N$, corresponding to the Majorana basis for $\mathfrak{su}(2^N)$. 

The value $f(i)$ is selected uniformly from the interval $[min,max]$ between the smallest $min$ and largest $max$ eigenvalues seen in our previous runs for \textbf{each} functional. In total, there were 7 functionals, four for $\mathfrak{su}(4)$ being $L_{24},L_{26}$ with $k=100,200$ and three for $\mathfrak{su}(8)$ being $L_{24}(k=1000)$ and $L_{26}$ with $k=500,1000$. For each of these 7 functionals, we have a $[min,max]$ interval, and we have 7 times 10,000 random functions in total.

Finally, we pick our values so that the condition $\text{det} = 1$ is enforced. To do so, we simply take the logs of $f(j)$ and turn the sampling problem into a convex body sampling problem for which there are many available methods and packages such as PyMC3 in Python. We should note that the problem of uniform sampling from the set of $f(j)$ satisfying those constraints is not exactly equivalent to the uniform sampling from the convex body that is formed by the $\log(f(j))$s, as we are making a change of coordinate by taking the logarithm. To the extent of our knowledge, packages such as PyMC3 can only be rigorously applied for convex body sampling. Nevertheless, from this study and side-experiments, we believe that the occurrence of local minima with exponential-like (Fermionic Brown-Susskind) growth are not random events but reflect a genuine propensity. Our experimental design does not enable us to claim this result with a precise confidence interval, for example the gold standard 5-sigmas, but we regard it as trustworthy.

For each random function, the loss for the best $L^2$-fit to the form $e^{\text{const. }(d-\frac{3.4^{N-1}N}{(4^N-1)})}$, const.\ $\geq 0$ was evaluated. How to compare these numbers with those of the actual runs? Assume the actual runs for a functional gave $r$ many distinct solutions (e.g. $r=5$ for $L_{24}(k=1000)$), with mean loss $l_{\text{mean}}$. We estimate the distribution of the mean loss of $r$ choices, i.e. $\frac{\sum_{i=1}^r l_{j_i}}{r}$, where $1\le j_1 < \ldots < j_r \le 10^4$, from the random losses $\{l_i\}_{i=1}^{10^4}$. Then, we can see if our $r$ actual losses are in general a rare $r$-sample of the random losses. 

\begin{figure}[h]
    \centering
    \includegraphics[scale = 0.5]{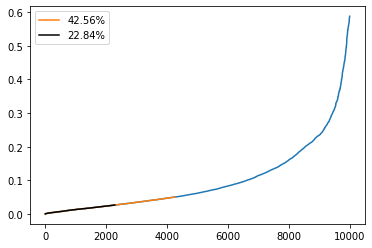}
    \includegraphics[scale = 0.5]{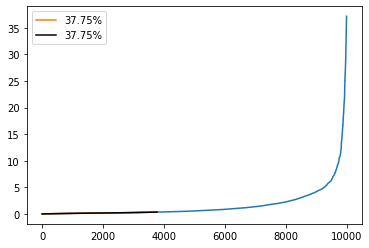}
    \includegraphics[scale = 0.5]{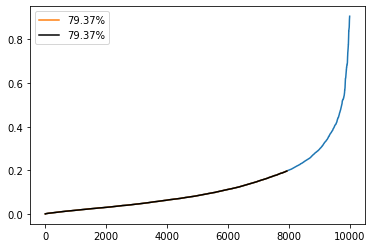}
    \includegraphics[scale = 0.5]{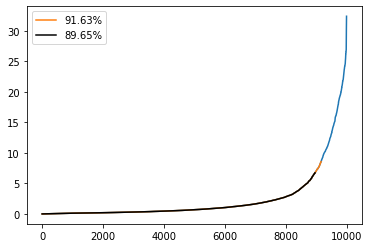}
    \caption{From top to bottom, left to right, we have plotted the sorted 10,000 random losses for each seven functionals: $L_{24}(k=100),L_{24}(k=200),L_{26}(k=100),L_{26}(k=200)$. }
    \label{randomlossesfig4}
\end{figure}

\begin{figure}[h]
    \centering
    \includegraphics[scale = 0.5]{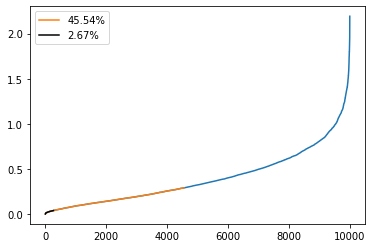}
    \includegraphics[scale = 0.5]{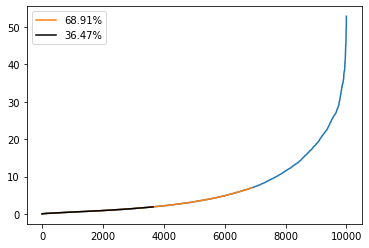}
    \includegraphics[scale = 0.5]{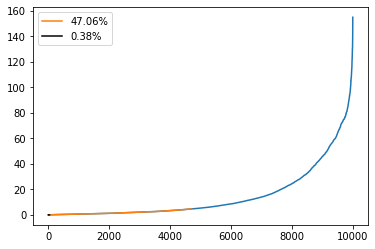}
    \caption{From top to bottom, left to right, we have plotted the sorted 10,000 random losses for each seven functionals: $L_{26}(k=500),L_{26}(k=1000), L_{24}(k=1000)$.}
    \label{randomlossesfig8}
\end{figure}

In \cref{randomlossesfig4} and \cref{randomlossesfig8}, we show how many of the random losses are less than $l_{mean}$, which is different than the $r$-sampling problem just discussed. In the plots, the (orange) black color shows the percentage of the random losses that are less than the (\textbf{mean} loss $l_{mean}$) \textbf{minimum} loss of the actual runs for that functional. We include mean loss for completeness, but we believe the number the most relevant to the argument we are making is the smallest loss, since it is not \textit{all} local minima, but rather \textit{some} O(1) fraction of them, which exhibit a close FBS structure

We sample $100$ million mean loss of subsets of size $r$ from the 10,000 losses for each of the 7 functionals. For each functional, we list the percentage of the samples that were lower than the actual $l_{mean}$ and the number of distinct solutions found: \begin{itemize}
    \item $\mathfrak{su}(4)$ functionals: $L_{24}(k=100): (21.92\%,3), L_{24}(k=200): (37.75\%,1), L_{26}(k=100): (79.37\%,1), L_{26}(k=200): (92.66\%,2)$,
    \item $\mathfrak{su}(8)$ functionals: $L_{24}(k=1000): (12.39\%,5), L_{26}(k=500): (29.93\%,3), L_{26}(k=1000): (59.45\%,5)$.
\end{itemize}

In summary, we conclude that the graphs in \cref{expsu4} and \cref{expsu8}, for $\mathfrak{su}(4)$ derived from local minima of the functional $L_{24}(k = 200)$, and for $\mathfrak{su}(8)$, derived from the local minima of the functional $L_{24}(k = 1000)$, exhibit a close Fermionic Brown-Susskind structure. We see a clear contrast between the $L_{24}$ and $L_{26}$ type functionals in this case, with only the former having local minima exhibiting a close FBS structure. This is the clearest distinction between the functional types $L_{24}$ (defined with an ``imaginary time'' exponential $e^{-k\ldots}$) and $L_{26}$ (defined with a ``Real time'' exponential $e^{ik \ldots}$) yet observed (referring back to \cite{freedman2021universe,freedman2021universeII}), and provides important feedback on the class of symmetry breaking functionals to be considered in future work. 

\section*{Acknowledgments}
The first named author would like to thank the Aspen Center for Physics for their continued hospitality. The second named author would like to acknowledge the support of the Perimeter Institute for Theoretical Physics and Microsoft. Research at Perimeter Institute is supported by the Government of Canada through Innovation, Science and Economic Development Canada and by the Province of Ontario through the Ministry of Research, Innovation and Science. The experiments were conducted using Microsoft computational resources.

\bibliographystyle{apa}
\bibliography{main}

\bigskip
\address{\textsuperscript{*\label{1}}
	Microsoft Research, Station Q, and Department of Mathematics, University of California, Santa Barbara, CA 93106, USA 
}

\address{\textsuperscript{$\dagger$\label{2}}
	Perimeter Institute for Theoretical Physics, Waterloo, ON N2L 2Y5, Canada \\
	\indent \textsuperscript{$\bullet$\label{3}} Research Consultant, Microsoft
}
\end{document}